\documentclass{ws-procs9x6-cpt25}
\begin{document}
	
	\providecommand{\textfrc}[1]{#1}
	
	\def\al{\alpha}
	\def\be{\beta}
	\def\ga{\gamma}
	\def\de{\delta}
	\def\ep{\epsilon}
	\def\ve{\varepsilon}
	\def\ze{\zeta}
	\def\et{\eta}
	\def\th{\theta}
	\def\vt{\vartheta}
	\def\io{\iota}
	\def\vka{\varkappa}
	\def\ka{\kappa}
	\def\la{\lambda}
	\def\vpi{\varpi}
	\def\rh{\rho}
	\def\vr{\varrho}
	\def\si{\sigma}
	\def\vs{\varsigma}
	\def\ta{\tau}
	\def\up{\upsilon}
	\def\ph{\phi}
	\def\vp{\varphi}
	\def\ch{\chi}
	\def\ps{\psi}
	\def\om{\omega}
	\def\Ga{\Gamma}
	\def\De{\Delta}
	\def\Th{\Theta}
	\def\La{\Lambda}
	\def\Si{\Sigma}
	\def\Up{\Upsilon}
	\def\Ph{\Phi}
	\def\Ps{\Psi}
	\def\Om{\Omega}
	\def\cA{{\cal A}}
	\def\cB{{\cal B}}
	\def\cC{{\cal C}}
	\def\cD{{\cal D}}
	\def\cE{{\cal E}}
	\def\cl{{\cal L}}
	\def\cL{{\cal L}}
	\def\cO{{\cal O}}
	\def\cV{{\cal V}}
	\def\cP{{\cal P}}
	\def\cR{{\cal R}}
	\def\cS{{\cal S}}
	\def\cT{{\cal T}}
	\def\mn{{\mu\nu}}
	
	\def\fr#1#2{{{#1}\over{#2}}}
	\def\frac#1#2{{\textstyle{{#1}\over{#2}}}}
	\def\half{{\textstyle{1\over 2}}}
	\def\ol{\overline}
	\def\prt{\partial}
	\def\pt{\phantom}
	
	\def\Re{\hbox{Re}\,}
	\def\Im{\hbox{Im}\,}
	
	\def\lsim{\mathrel{\rlap{\lower4pt\hbox{\hskip1pt$\sim$}}
			\raise1pt\hbox{$<$}}}
	\def\gsim{\mathrel{\rlap{\lower4pt\hbox{\hskip1pt$\sim$}}
			\raise1pt\hbox{$>$}}}
	
	\def\etal{{\it et al.}}
	
	\def\vev#1{\langle {#1}\rangle}
	\def\expect#1{\langle{#1}\rangle}
	\def\bra#1{\langle{#1}|}
	\def\ket#1{|{#1}\rangle}
	
	\def\tr{{\rm tr}}
	
	\def\eff{{\rm eff}}
	
	\def\sqr#1#2{{\vcenter{\vbox{\hrule height.#2pt
					\hbox{\vrule width.#2pt height#1pt \kern#1pt
						\vrule width.#2pt}
					\hrule height.#2pt}}}}
	
	\newcommand{\beq}{\begin{equation}}
		\newcommand{\eeq}{\end{equation}}
	\newcommand{\bea}{\begin{eqnarray}}
		\newcommand{\eea}{\end{eqnarray}}
	\newcommand{\rf}[1]{(\ref{#1})}
	
	\newcommand{\bM}{\begin{pmatrix}}
		\newcommand{\eM}{\end{pmatrix}}
	
	\def\nn{\nonumber}
	
	\def\w{w}
	\def\f{w}
	
	\def\Psb{\bar{\Ps}}
	\def\psb{\bar{\ps}}
	\def\psfb{\bar{\ps}_{\f}}
	
	\def\bsi{\ol\si}
	\def\Omb{\widebar\Om}
	
	\def\opr#1{\overrightarrow #1}
	\def\opl#1{\overleftarrow #1}
	\def\oplr#1{\overleftrightarrow #1}
	
	\def\mbf#1{\boldsymbol #1}
	
	\def\syjm#1#2{{}_{#1}Y_{#2}}
	
	\def\O{\mathcal O}
	\def\Q{\mathcal Q}
	\def\S{\mathcal S}
	\def\P{\mathcal P}
	\def\V{\mathcal V}
	\def\A{\mathcal A}
	\def\T{\mathcal T}
	\def\C{\mathcal C}
	\def\G{\mathcal G}
	\def\E{\mathcal E}
	\def\B{\mathcal B}
	\def\K{\mathcal K}
	\def\R{\mathcal R}
	
	\def\cEvec{\mbf\E}
	\def\cBvec{\mbf\B}
	
	\def\Avec{\mbf A}
	\def\Evec{\mbf E}
	\def\Bvec{\mbf B}
	\def\pvec{\mbf p}
	\def\kvec{\mbf k}
	\def\nv{n}
	\def\nvec{\mbf \nv}
	\def\vvec{\mbf v}
	\def\xvec{\mbf x}
	\def\Pvec{\mbf P}
	\def\gavec{\mbf\ga}
	\def\thvec{\mbf\th}
	\def\sivec{\mbf\si}
	
	\def\del{\mbf\nabla}
	
	\def\Jvec{\mbf J}
	\def\Lvec{\mbf L}
	\def\Svec{\mbf S}
	\def\Kvec{\mbf K}
	\def\Mvec{\mbf M}
	\def\Tvec{\mbf T}
	\def\bevec{\mbf\be}
	\def\tavec{\mbf\ta}
	\def\pchvec{\widetilde{\mbf p}}
	\def\pch{\widetilde{p}}
	
	\def\pmag{|\pvec|}
	\def\komag{|{\mbf k_0}|}
	\def\krmag#1{|{\mbf k_r}(#1)|}
	
	\def\punit{\hat p}
	\def\kunit{\hat k}
	\def\xunit{\hat x}
	\def\yunit{\hat y}
	\def\zunit{\hat z}
	\def\epunit{\hat\ep}
	\def\thunit{\hat\th}
	\def\phunit{\hat\ph}
	
	\def\phat{\mbf\punit}
	\def\khat{\mbf\kunit}
	\def\xhat{\mbf\xunit}
	\def\yhat{\mbf\yunit}
	\def\zhat{\mbf\zunit}
	\def\ephat{\mbf\epunit}
	\def\thhat{\mbf\thunit}
	\def\phhat{\mbf\phunit}
	
	\def\opr#1{\overrightarrow #1}
	\def\opl#1{\overleftarrow #1}
	\def\oplr#1{\overleftrightarrow #1}
	
	\def\widebar{\overline}
	\def\bigvev#1{\big\langle #1 \big\rangle}
	\def\biggvev#1{\bigg\langle #1 \bigg\rangle}
	\def\free{{\rm free}}
	
	\def\cg#1#2{\langle #1 | #2 \rangle}
	\def\bc#1#2{\left(\begin{smallmatrix} #1 \\ #2 \end{smallmatrix}\right)}
	
	\def\Qhat{\widehat\Q}
	\def\Shat{\widehat\S}
	\def\Phat{\widehat\P}
	\def\Vhat{\widehat\V}
	\def\Ahat{\widehat\A}
	\def\That{\widehat\T}
	\def\Tdual{\widetilde{\widehat\T}\phantom{}}
	\def\Chat{\widehat\C}
	\def\Ghat{\widehat\G}
	\def\Gahat{\widehat\Ga}
	\def\Mhat{\widehat M}
	
	\def\mhat{\widehat m}
	\def\mfivehat{\widehat m_5}
	\def\ahat{\widehat a}
	\def\bhat{\widehat b}
	\def\chat{\widehat c}
	\def\dhat{\widehat d}
	\def\ehat{\widehat e}
	\def\fhat{\widehat f}
	\def\ghat{\widehat g}
	\def\Hhat{\widehat H}
	\def\gdual{\widetilde{\widehat g}\phantom{}}
	\def\Hdual{\widetilde{\widehat H}\phantom{}}
	\def\gt{\widetilde g}
	\def\Ht{\widetilde H}
	
	\def\Htf#1#2#3{{{\tilde{H}}_{#1}}\,\hspace{-1 pt}^{(#2)#3}_\eff}
	\def\gtf#1#2#3{{{\tilde{g}}_{#1}}\,\hspace{-1 pt}^{(#2)#3}_\eff}
	
	\def\X{X}
	\def\Y{Y}
	\def\Z{Z}
	\def\Xhat{\widehat\X}
	\def\Yhat{\widehat\Y}
	\def\Zhat{\widehat\Z}
	
	\def\template#1#2#3#4{#1^{(#2)#4}_{#3}}
	\def\acoef#1#2{\template{a}{#1}{#2}{}}
	\def\bcoef#1#2{\template{b}{#1}{#2}{}}
	\def\ccoef#1#2{\template{c}{#1}{#2}{}}
	\def\dcoef#1#2{\template{d}{#1}{#2}{}}
	\def\gzBcoef#1#2{\template{g}{#1}{#2}{(0B)}}
	\def\goBcoef#1#2{\template{g}{#1}{#2}{(1B)}}
	\def\goEcoef#1#2{\template{g}{#1}{#2}{(1E)}}
	\def\HzBcoef#1#2{\template{H}{#1}{#2}{(0B)}}
	\def\HoBcoef#1#2{\template{H}{#1}{#2}{(1B)}}
	\def\HoEcoef#1#2{\template{H}{#1}{#2}{(1E)}}
	\def\nr{{\rm NR}}
	
	\def\nrtemplate#1#2#3{#1^{\nr#3}_{#2}}
	
	\def\cs133{\rm Cs}
	
	\def\anr#1{\nrtemplate{a}{#1}{}}
	\def\bnr#1{\nrtemplate{b}{#1}{}}
	\def\bbnr#1{\nrtemplate{b}{#1}{'}}
	\def\cnr#1{\nrtemplate{c}{#1}{}}
	\def\dnr#1{\nrtemplate{d}{#1}{}}
	\def\ddnr#1{\nrtemplate{d}{#1}{'}}
	\def\gzBnr#1{\nrtemplate{g}{#1}{(0B)}}
	\def\goBnr#1{\nrtemplate{g}{#1}{(1B)}}
	\def\goEnr#1{\nrtemplate{g}{#1}{(1E)}}
	\def\HzBnr#1{\nrtemplate{H}{#1}{(0B)}}
	\def\HoBnr#1{\nrtemplate{H}{#1}{(1B)}}
	\def\HoEnr#1{\nrtemplate{H}{#1}{(1E)}}
	
	\def\Vnr#1{\nrtemplate{\V}{#1}{}}
	\def\TzBnr#1{\nrtemplate{\T}{#1}{(0B)}}
	\def\ToBnr#1{\nrtemplate{\T}{#1}{(1B)}}
	\def\ToEnr#1{\nrtemplate{\T}{#1}{(1E)}}
	\def\Vnrf#1#2{\nrtemplate{{\V_{#1}}}{#2}{}}
	\def\TzBnrf#1#2{\nrtemplate{{\T_{#1}}}{#2}{(0B)}}
	\def\ToEnrf#1#2{\nrtemplate{{\T_{#1}}}{#2}{(1E)}}
	\def\ToBnrf#1#2{\nrtemplate{{\T_{#1}}}{#2}{(1B)}}
	
	\def\anrf#1#2{\nrtemplate{{a_{#1}}}{#2}{}}
	\def\cnrf#1#2{\nrtemplate{{c_{#1}}}{#2}{}}
	\def\gzBnrf#1#2{\nrtemplate{{g_{#1}}}{#2}{(0B)}}
	\def\goBnrf#1#2{\nrtemplate{{g_{#1}}}{#2}{(1B)}}
	\def\goEnrf#1#2{\nrtemplate{{g_{#1}}}{#2}{(1E)}}
	\def\HzBnrf#1#2{\nrtemplate{{H_{#1}}}{#2}{(0B)}}
	\def\HoBnrf#1#2{\nrtemplate{{H_{#1}}}{#2}{(1B)}}
	\def\HoEnrf#1#2{\nrtemplate{{H_{#1}}}{#2}{(1E)}}
	
	\def\sVnrf#1#2{\nrtemplate{{\V_{#1}}}{#2}{,{\rm Sun}}}
	\def\sanrf#1#2{\nrtemplate{{a_{#1}}}{#2}{,{\rm Sun}}}
	\def\scnrf#1#2{\nrtemplate{{c_{#1}}}{#2}{,{\rm Sun}}}
	\def\sgzBnrf#1#2{\nrtemplate{{g_{#1}}}{#2}{(0B), {\rm Sun}}}
	\def\sgoBnrf#1#2{\nrtemplate{{g_{#1}}}{#2}{(1B),{\rm Sun}}}
	\def\sHzBnrf#1#2{\nrtemplate{{H_{#1}}}{#2}{(0B),{\rm Sun}}}
	\def\sHoBnrf#1#2{\nrtemplate{{H_{#1}}}{#2}{(1B),\rm Sun}}
	\def\sTzBnrf#1#2{\nrtemplate{{\T_{#1}}}{#2}{(0B),{\rm Sun}}}
	\def\sToBnrf#1#2{\nrtemplate{{\T_{#1}}}{#2}{(1B),\rm Sun}}
	
	\def\m{m_\ps}
	
	\def\widecheck#1{\hskip#1pt\huge$\check{}$}
	\def\bighacek#1#2{\vbox{\ialign{##\crcr\widecheck#2\crcr
				\noalign{\kern-9.5pt\nointerlineskip}
				$\hfil\displaystyle{#1}\hfil$\crcr}}}
	\def\hb{\bighacek{b}{2}}
	\def\hd{\bighacek{d}{2}}
	\def\hg{\bighacek{g}{3}}
	\def\hh{\bighacek{H}{5}{}}
	\def\hH{\bighacek{H}{5}{}}
	\def\hk{\bighacek{\K}{5}{}}
	
	\def\mr{\ol{m}_{\rm r}}
	\def\xm{\xi}
	
	\def\codt{\cos{\om_\oplus T_\oplus}}
	\def\sodt{\sin{\om_\oplus T_\oplus}}
	\def\ctodt{\cos{2\om_\oplus T_\oplus}}
	\def\stodt{\sin{2\om_\oplus T_\oplus}}
	\def\ce{\cos\et}
	\def\se{\sin\et}
	\def\cc{\cos\ch}
	\def\sc{\sin\ch}
	\def\cto{\cos{\Om_\oplus T}}
	\def\sto{\sin{\Om_\oplus T}}
	\def\gai{\fr 1 \ga}
	\def\gait{\fr 1 {2\ga}}
	
	\def\gm#1#2#3{g^{(M)}_{#1#2#3}}
	\def\gA#1{g^{(A){#1}}}
	\def\glA#1{g^{(A)}_{#1}}
	\def\ggA#1#2#3{g^{(A)}_{#1#2#3}}
	\def\gT#1{g^{(T)}_{#1}}
	\def\guT#1{g^{(T)#1}}
	\def\ggT#1#2#3{g^{(T)}_{#1#2#3}}
	
	\def\k{k}
	\def\n{n}
	\def\q{q}
	\def\atm{{\rm atom}}
	\def\atmb{{\overline{\rm H}}}
	
	\def\AzB#1#2{{\La_{#1}}^{(0B)}_{#2}}
	\def\AoB#1#2{{\La_{#1}}^{(1B)}_{#2}}
	\def\AzE#1#2{{\La_{#1}}^{(0E)}_{#2}}
	\def\Agen#1{{\La_{#1}}^{(qP)}_{kj}}
	
	\def\chM{\vartheta}
	\def\phM{\varphi}
	\def\AM{A}
	
	\def\eb{{\overline{e}}}
	\def\pb{{\overline{p}}}
	\def\epb{{\overline{\ep}}}
	
	\def\Vb{{\overline{\mathcal V}}}
	\def\Tb{{\overline{\mathcal T}}}
	\def\ab{{\overline{a}}}
	\def\cb{{\overline{c}}}
	\def\fb{{\overline{\f}}}
	\def\gb{{\overline{g}}}
	\def\nub{{\overline{\nu}}}
	\def\AMb{{\overline{\AM}}}
	\def\Rb{{\overline{R}}}
	\def\Hb{{\overline{H}}}
	\def\Tb{{\overline{T}}}
	
	\def\ring#1{{\mathaccent'27 #1}}
	
	\def\nrfctemplate#1#2{\nrtemplate{\ring{#1}}{#2}{}}
	\def\anrfc#1{\nrfctemplate{a}{#1}}
	\def\cnrfc#1{\nrfctemplate{c}{#1}}
	\def\Vnrfc#1{\nrfctemplate{\mathcal{V}}{#1}}
	
	\def\afbx#1{(\ab^{#1}_{\rm{eff}})}
	\def\cbx#1{(\cb^{#1})}
	\def\cbw{\cbx{w}}
	\def\afbw{\afbx{w}}
	\def\mt{m^{\rm T}}
	\def\ms{m^{\rm S}}
	\def\mb{m^{\rm{B}}}
	
	\def\fctemplate#1#2#3{\ring{#1}^{(#2)}_{#3}}
	\def\afc#1#2{\fctemplate{a}{#1}{#2}}
	\def\cfc#1#2{\fctemplate{c}{#1}{#2}}
	
	\def\afcb#1#2{\fctemplate{\ab}{#1}{#2}}
	\def\cfcb#1#2{\fctemplate{\cb}{#1}{#2}}
	\newcommand{\mathfrc}[1]{\text{\textfrc{#1}}}
	\def\pd{\ol{p}\ol{d}}
	\def \j{j}
	\def\ii{a}
	\def\AF{A}
	\def\Ce#1#2{C^{#1}_{#2}}
	\def\Ae#1#2{W^{#1}_{#2}}
	\def\AI{W^I}
	\def\AJ{W^J}
	\def\CJ#1#2{C^{#1}_{#2}}
	\def\CI#1#2{C^{#1}_{#2}}
	\def\eiso{\de\ring{\ep}}
	\def\LaF#1#2{\La_{#1}^{#2}}
	\def\Mco#1#2{M_{#1}^{#2}}
	\def\alg{\al'}
	\def\Ht{\widetilde{H}}
	\def\gt{\widetilde{g}}
	\def\Va#1#2#3{V_{#1}^{(#2)#3}}
	\def\Tg#1#2#3{T_{#1}^{(#2)#3}}
	\def\dC{\la}
	\def\caz{\rm Ca}
	\def\Jj#1{{J_#1}}
	
	\def\Bunit{\hat B}
	\def\Bhat{\mbf\Bunit}
	
	\def\TL{T_L}
	
	\def\TqBnr#1{\nrtemplate{\T}{#1}{(qB)}}
	
	\def\bVnrf#1#2{\nrtemplate{{\widebar{\V}_{#1}}}{#2}{}}
	
	\title{Nonminimal Lorentz Violation in Atomic and Molecular Spectroscopy Experiments}
	
	\author{Arnaldo J.\ Vargas}
	
	\address{Laboratory of Theoretical Physics, Department of Physics, University of Puerto Rico,\\
		R\'{\i}o Piedras 00936, Puerto Rico}
	
	\begin{abstract}
		This presentation discusses potential signals of Lorentz violation that could be observed in atomic and molecular spectroscopy experiments. It provides a general overview of the nonrelativistic effective SME coefficients that have been constrained, as well as the prospects for placing first-time bounds on those that remain unconstrained. Additionally, it highlights the importance of considering transitions involving atomic or molecular states with high angular momentum in tests of Lorentz symmetry.
	\end{abstract}
	
	\bodymatter
	
	\section{Introduction}
	
	The possibility that tiny deviations from Lorentz symmetry could serve as low-energy signals of physics beyond the Standard Model and General Relativity\cite{ksp} led to the development of the Standard-Model Extension (SME),\cite{ck} a comprehensive framework for systematically testing Lorentz symmetry. The SME assigns a unique coefficient for Lorentz violation, known as an SME coefficient, to each Lorentz-violating operator. The goal of the SME is to search for evidence of a nonzero SME coefficient, which would indicate a violation of Lorentz symmetry.
	
	The SME also provides a framework for systematically testing CPT symmetry, since CPT violation implies Lorentz violation in a realistic effective quantum field theory.\cite{owg} Approximately half of the Lorentz-violating operators in the SME also violate CPT symmetry. The coefficients associated with these operators are referred to as coefficients for CPT violation, and any evidence of a nonzero value for such a coefficient would suggest CPT violation.
	
	Experiments that probe these symmetries are typically sensitive to a limited subset of SME coefficients. Therefore, a wide range of experiments is necessary for the systematic testing of Lorentz and CPT symmetry. Using the SME framework, one can quantify the sensitivity of an experiment to specific SME coefficients, which allows for comparing the competitiveness and complementarity of different tests of these symmetries.
	
	Early versions of the SME focused on Lorentz-violating operators of mass dimension $d \leq 4$, referred to as minimal operators, to avoid introducing nonrenormalizable terms into the framework.\cite{ck} However, this restriction is not essential, since the SME is based on the premise that any underlying Lorentz-violating theory would manifest as a low-energy effective field theory. Such a theory does not need to be renormalizable if it is only valid within a limited energy range. Consequently, there is no reason to exclude nonminimal operators, defined as those with $d \geq 5$, from the SME. Subsequent extensions of the SME have included these nonminimal operators.\cite{km09,km12,km13,nonmingrav,kl21} In some ways, the nonminimal operators are more interesting than the minimal ones, as their effects are expected to be suppressed by the ratio of the experimental energy scale to the high-energy scale at which the effective theory breaks down, providing a plausible explanation for the tiny effects, currently unobserved, of Lorentz and CPT violation.
	
	The following presents an overview of models for testing Lorentz and CPT symmetry in atomic and molecular spectroscopy experiments that incorporate nonminimal effects, with particular emphasis on those developed in Refs.\ \refcite{gkv14,kv15,kv18,va24,va25,mu25,va25a}. Owing to the focused scope of this discussion, we do not consider high-precision spectroscopy experiments that test models involving only minimal SME coefficients and have not been used to impose bounds on nonminimal ones.\cite{minspec}
	
	In many cases, the contributions from SME coefficients to the observables of a system appear in specific linear combinations. For convenience, these linear combinations are defined as effective SME coefficients. In the model discussed here, the effective coefficients are the nonrelativistic (NR) coefficients.\cite{km13}
	
	The structure of this presentation is as follows. We provide a brief description of the NR coefficients and their relevant properties. We then discuss potential signals of Lorentz violation that could be observed in atomic and molecular experiments. Also, we review the NR coefficients that have already been constrained and highlight potential experiments that could impose bounds on the remaining unconstrained NR coefficients.
	
	\section{Main}
	
	The dominant Lorentz-violating effects on the atomic spectrum can be obtained by considering only the leading Lorentz-violating perturbation to the free propagation of the fermions within the atom.\cite{kv15} The perturbation of interest to us is the one obtained by adding a single-particle perturbation derived in Ref.\ \refcite{km13}, containing both minimal and nonminimal operators, for each fermion in the atom. The perturbation depends on the free-particle energy $\sqrt{\pmag^2 + m_w^2}$, which can be expanded in powers of $\pmag/m_w$ if it is smaller than one. The resulting perturbation, obtained by applying this expansion to the free-particle energy, is called the nonrelativistic (\nr) perturbation, denoted by $\delta h^\nr_w$. Here, $w$ denotes the fermion type and takes the values $e$ for electron, $p$ for proton, and $\mu$ for muon.
	
	As demonstrated in Ref.\ \refcite{kv15}, some terms in the single-particle perturbation do not contribute to the energy of atomic or molecular levels. After excluding these terms, the Lorentz-violating single-particle perturbation takes the form
	\bea
	\de h_\f^\nr&=& -\sum_{k \j m} \pmag^k
	\syjm{0}{\j m}(\phat)
	\left(\Vnrf{\f}{\k \j m}+\sivec\cdot\ephat_r\TzBnrf{\f}{k \j m}\right)\nn\\
	&&+\sum_{k \j m} \pmag^k\sivec\cdot\left(\syjm{+ 1}{\j m}(\phat) \ephat_--\syjm{- 1}{\j m}(\phat) \ephat_+\right)
	\ToBnrf{\f}{k \j m},
	\label{nr}
	\eea
	where $k$ takes the values $0$, $2$, and $4$ with $0\leq j \leq 5$ and $-j\leq m \leq j$.
	
	The use of the nonrelativistic expansion of the free-particle energy leads to linear combinations of SME coefficients of different mass dimensions in the perturbation, each accompanied by appropriate powers of the fermion mass. These linear combinations of coefficients are called the nonrelativistic (NR) coefficients and are distinguished by the superscript NR. The relation between the NR coefficients and the standard SME coefficients for Lorentz and CPT violation is provided in Eqs.\ (111) and (112) of Ref.\ \refcite{km13}. In general, all NR coefficients receive contributions from nonminimal SME coefficients, while some also receive contributions from minimal ones.
	
	The NR coefficients appearing in $\delta h_w^\nr$, namely $\Vnrf{\f}{k \j m}$, ${\T_\f}^{\nr(0B)}_{k\j m}$ and ${\T_\f}^{\nr(1B)}_{k\j m}$, are linear combinations of SME coefficients associated with operators with a definite CPT sign. These combinations are
	\bea
	\Vnrf{\f}{k\j m} &=& \cnrf{\f}{k\j m} - \anrf{\f}{k\j m}, \nn \\ {\T_\f}^{\nr(qP)}{k\j m} &=& {g_\f}^{\nr(qP)}{k\j m} - {H_\f}^{\nr(qP)}_{k\j m},
	\label{cpt}
	\eea
	where the $a$- and $g$-type coefficients correspond to CPT-odd operators, while the $c$- and $H$-type coefficients are associated with CPT-even ones. In contrast, nonrelativistic antimatter experiments are sensitive to a different linear combination. To distinguish between these two scenarios, the index $\fb$ is used to denote the antiparticle case, in contrast to $w$, which refers to the particle one. The combinations relevant to antimatter experiments are given by
	\bea
	\Vnrf{\fb}{k\j m} &=& \cnrf{\f}{k\j m} + \anrf{\f}{k\j m}, \nn \\
	{\T_\fb}^{\nr(qP)}{k\j m} &=& -{g_\f}^{\nr(qP)}{k\j m} - {H_\f}^{\nr(qP)}_{k\j m}.
	\label{cptb}
	\eea
	The difference between Eqs.~\eqref{cpt} and \eqref{cptb} lies in the sign of the CPT-odd terms. It is important to note that coefficients with definite CPT sign are always defined relative to the particle type, not the antiparticle one, which is consistent with standard conventions in the literature. Finally, the indices $j$ and $m$ of the NR coefficients correspond to those of the spin-weighted spherical harmonics $\syjm{s}{j m}(\phat)$ appearing in Eq.~\rf{nr}, while the index $k$ indicates the power of the fermion's momentum magnitude $\pmag$.
	
	Not all combinations of the $k$ and $j$ indices are allowed in the perturbation \rf{nr}. For the $\V$-type coefficients, only even values of $j$ are permitted, and $k$ must satisfy $k \geq j$. For the $\T$-type coefficients, $j$ is odd and must satisfy $k + 1 \geq j$. Overall, the model based on the perturbation \rf{nr} contains 178 nonrelativistic (NR) coefficients accessible in atomic spectroscopy experiments, per particle type. For example, hydrogen spectroscopy experiments are sensitive to 178 electron coefficients and an equal number of proton coefficients.
	
	It is important to note that not all 178 coefficients contribute to the Lorentz-violating energy shift for a given atomic level, as discussed in Ref.\ \refcite{kv15}. For instance, in two-fermion atoms such as hydrogen, antihydrogen, and muonium, only coefficients with index $j$ satisfying $j \leq 2J$ and $j \leq 2F$ can contribute to a level characterized by quantum number $J$ (the total angular momentum of the lighter fermion) and $F$ (the total atomic angular momentum).\cite{kv15} For instance, transitions between $nS$ states in two-fermion atoms with quantum numbers $J = 1/2$ and $F \leq 1$ are sensitive to only 40 coefficients per particle type---a small fraction of the total number of coefficients included in the model.
	
	If rotational symmetry is broken, the atomic spectrum may depend on the orientation of the total angular momentum $\vec{F}$ of the atom. This could lead to unexpected multiple resonance peaks or broadening of the line shape.\cite{gkv14,kv15,mu25} However, the most common experimental signal of rotational symmetry violation in Earth-based experiments is a sidereal variation of the resonant frequency. The basic idea is that the atomic spectrum changes when the atom, specifically its total angular momentum $\vec{F}$, is rotated relative to a fixed inertial reference frame. This rotation can be achieved by applying an external magnetic field that is adiabatically rotated with respect to a fixed frame. A signal of rotational symmetry breaking would then manifest as a time variation in the resonance frequencies of the atom, modulated at the rotation frequency of the magnetic field. For experiments conducted on the surface of the Earth, it is sufficient to fix the magnetic field in the laboratory frame, since the Earth's rotation naturally produces an adiabatic rotation relative to a fixed inertial frame. In this case, the modulation frequency corresponds to the sidereal frequency.
	
	The SME coefficients transform under observer transformations; therefore, their values depend on the reference frame used to analyze the experiment. To enable comparison across multiple experiments, it is standard practice to report constraints on the coefficients in a common frame, typically the Sun-centered frame.\cite{sunframe,tables}
	The indices $j$ and $m$ of the NR coefficients in the Sun-centered frame that contribute to the Lorentz-violating shift in the resonant frequency provide insight into the type of sidereal variation expected in an experiment. Suppose the time variation of the transition frequency is decomposed into harmonics of the sidereal frequency, $\omega_\oplus \simeq 2\pi/(1436\, {\rm min})$. An NR coefficient in the Sun-centered frame with index $m$ can contribute only to the amplitude of the $|m|$-th harmonic of $\omega_\oplus$.\cite{kv15} Since $|m| \leq j$, a coefficient with index $j$ cannot contribute to harmonics higher than the $j$-th harmonic of the sidereal frequency.
	
	The NR coefficients with $j \leq 2$ contain the dominant contributions from the minimal SME coefficients, specifically the $\T$-type with $k = 0$ for $j = 1$ and the $\V$-type with $k = 2$ for $j = 0$ and $j=2$. For this reason, many experiments motivated by the minimal SME focused on these types of coefficients, and the constraints obtained from those experiments have since been reinterpreted as bounds on the NR coefficients with $j = 1$ and $j = 2$. In particular, studies of sidereal variation in resonance frequencies within the ground state of muonium\cite{mu01} and hydrogen,\cite{maser} as well as experiments using a Xe-He co-magnetometer,\cite{xema} have been used to place bounds on the $j = 1$ $\T$-type NR coefficients with $|m| = 1$ in the electron, proton, muon, and neutron sectors of the SME.\cite{gkv14,kv15,kv18} A sidereal variation study using a Ne-Rb-K co-magnetometer,\cite{NeRb} which investigated variations at the first two harmonics of the sidereal frequency, was used to constrain the $j = 2$ $\V$-type NR coefficients for the neutron, with $|m| = 1$ and $|m| = 2$.\cite{kv18}
	
	NR coefficients with index $j \geq 3$ remain unconstrained,\cite{tables} as all transitions considered thus far have involved states with total angular momentum less than or equal to $F = 1$. Consequently, Lorentz and CPT tests involving transitions to states with higher angular momentum could offer sensitivity to these unconstrained NR coefficients. For example, the proposed measurement of the $2S$--$2P_{3/2}$ transition in muonium could provide the first bounds on $j = 3$ muon NR coefficients\cite{mu25} and sidereal variation studies of the rovibrational transitions\cite{h2+} within the ground state of H$_2^+$ could impose the first bounds on the proton NR coefficients with $j = 4$.\cite{va25a}
	
	Another crucial feature of the perturbation \rf{nr} is its dependence on the magnitude of the momentum, which allows for some systems to exhibit significantly greater sensitivity to certain coefficients, particularly those with $k = 4$, due to the higher momentum of their constituents. For example, the proton in deuterium has a higher momentum than in hydrogen because of its motion within the deuteron. Consequently, deuterium can be significantly more sensitive to proton coefficients than hydrogen, even though hydrogen typically offers better sensitivity to Lorentz-violating frequency shifts.\cite{kv15,va24} A sidereal variation study of the Zeeman-hyperfine transitions within the ground state of deuterium, currently in its final stages of analysis, is expected to improve bounds on the proton NR coefficients with $j = 1$ for $k = 2$ and $k = 4$, and to provide the first constraints on the proton NR coefficients with $j = 2$.\cite{va24} Another system benefiting from this momentum enhancement is muonic hydrogen, in which the negative muon has greater momentum than the positive muon in muonium. Therefore, although the proposed improved measurement of the hyperfine splitting in the ground state of muonium\cite{museum} is the best candidate for improving bounds on the $\T$-type $k = 0$ muon coefficients, the best prospects for tightening bounds on the $k = 2$ and $k = 4$ muon coefficients lie in sidereal variation studies using the data collected in muonic hydrogen experiments,\cite{gkv14} such as the one currently underway.\cite{muH} Finally, rovibrational transitions within the ground state of HD$^+$ have also been measured experimentally.\cite{hd+} Owing to the higher momentum of the proton within the deuteron, these transitions may offer enhanced sensitivity to proton coefficients compared to those in H$_2^+$.\cite{va25a}
	
	Nonrelativistic (NR) coefficients in the Sun-centered frame with $m = 0$ do not induce sidereal variations in the resonance frequency.\cite{gkv14,kv15,kv18} However, these coefficients can still lead to constant frequency shifts, which can be constrained through alternative methods. One such method exploits the dependence of certain $m = 0$ coefficients on the relative orientation between the magnetic field and Earth's rotation axis.\cite{kv15} An experiment studying transitions within the ground state of hydrogen for different magnetic field orientations was able to place limits on the $\T$-type proton and electron coefficients with $j = 1$, $m = 0$.\cite{no24} Another strategy for probing $m = 0$ coefficients involves tests of CPT symmetry.\cite{kv15} For example, comparisons between the $1S$--$2S$ transitions in hydrogen and antihydrogen\cite{ah18} have yielded bounds on the $a$-type coefficients with $j = 0$ for both electrons and protons.\cite{kv18} Additionally, coefficients with $m = 0$ can be constrained by comparing precise experimental results with predictions from the Standard Model. This approach has led to bounds on $a$-type and $c$-type coefficients in the electron, proton, and muon sectors. Specific examples include the $1S$--$2S$ transition in positronium,\cite{kv15} the $1S$--$2S$ transition and classical Lamb shift in muonium,\cite{gkv14,oh22} and the $1S$--$2P$ transition in antihydrogen.\cite{va25,1s2p} As with sidereal variation studies, it is desirable to investigate transitions involving states with higher angular momentum quantum numbers. In this context, the proposed possibility of performing molecular antihydrogen spectroscopy\cite{h2+,hbar2} is particularly exciting, as it offers sensitivity to NR coefficients with $j \geq 3$ and $m = 0$, which remain inaccessible in the transitions studied thus far in antihydrogen.\cite{kv15,va25a}
	
	\section{Final Remarks}
	
	Among the NR coefficients considered in the models discussed here, only about $16\%$ have been constrained in the muon sector and roughly $25\%$ in the electron, proton, and neutron sectors. This leaves substantial room for exploring possible violations of CPT and Lorentz symmetry in atomic and molecular experiments. A key reason for the large number of unconstrained NR coefficients is the scarcity of Lorentz and CPT tests involving transitions with high angular momentum states. While it remains important to continue tightening existing bounds, many of which can still be significantly improved, it is equally important to design and carry out experiments that specifically probe the currently unconstrained coefficients.

\end{document}